# Real-Time Go-Around Prediction
## A case study of JFK airport

Kaijing Ding[1], Ke Liu[1], Lu Dai, Mark Hansen
Department of Civil and Environmental Engineering
University of California, Berkeley
Berkeley, United States
kaijing@berkeley.edu, liuke126@berkeley.edu,
dailu@berkeley.edu, mhansen@ce.berkeley.edu

Kennis Chan, John Schade
ATAC Corporation
Santa Clara, United States
koc@atac.com, jes@atac.com

*Abstract*— **In this paper, we employ the long-short term memory model (LSTM) to predict the real-time go-around probability as an arrival flight is approaching JFK airport and within 10 nm of the landing runway threshold. We further develop methods to examine the causes to go-around occurrences both from a global view and an individual flight perspective. According to our results, in-trail spacing, and simultaneous runway operation appear to be the top factors that contribute to overall go-around occurrences. We then integrate these pre-trained models and analyses with real-time data streaming, and finally develop a demo web-based user interface that integrates the different components designed previously into a real-time tool that can eventually be used by flight crews and other line personnel to identify situations in which there is a high risk of a go-around.**

I. INTRODUCTION

A go-around is an aborted landing of an aircraft during its final approach or after it has already touched down. This action can be either initiated by the pilot flying or requested by air traffic control due to various reasons, such as an unstable approach or a runway obstruction. While a go-around is implemented as a precautionary measure to enhance flight safety, it poses challenges as it disrupts normal airport operations, increases the workload for air traffic controllers and flight crews [1], elevates noise levels and fuel consumption [2], and impacts airport throughput [3] and flight on-time performance [4].

Despite the high risks and costs associated with go-around occurrence, the decision-making on go-around initiation varies from airline to airline and from person to person. Typically, airline procedures state that if a pilot determines the approach is unstable at specified altitudes according to the airline established policies, they should go around. However, interviews suggest that the decisions on go-arounds are strongly influenced by individual experiences and mental states [4]. As a result, "the collective industry performance of complying with go-around policies is extremely poor and only about 3% of unstable approaches result in a go-around" [5]. Multiple studies have attempted to examine today's stabilized approach criteria and develop detailed go-around criteria based on interviews and flight simulator experiments [5-8].

Meanwhile, go-arounds have appeared more frequently in the recent years. After the pandemic (from FY17-19 to FY21-22), the go-around occurrence rate as percent of arrivals at US Core 30 airport grew from 0.3% to 0.4%, with an increase of 24.9% [9]. The FAA report attributes this change to "mainly … a large increase in go-arounds relative to a smaller increase in arrival operations" while failing to further elaborate the reasons behind the "large increase in go-arounds." A New York Times article [10], investigated the abnormally high occurrence rate of airline close calls in 2023 and concluded that "mistakes by air traffic controllers — stretched thin by a nationwide staffing shortage — have been one major factor." Since go-arounds and close calls are the two possible results of a loss of separation during landing, we may speculate the same reason to account for the increasing go-around rate.

Considering the high risks and costs, the current highly subjective initiation criteria, and the growing occurrence rate of go-around, this study seeks to enable real-time go-around probability prediction informed by historical data. This study builds upon a series of previous work which develops the go-around detection algorithm and studies the underlying causes of go-around occurrence [11]; proposes the concept of runway occupancy buffer as a feature to predict go-around occurrence [12]; and predicts go-around occurrence with Input-Output Hidden Markov Model [13]. In this study, we will (1) provide a real-time go-around prediction tool based on various spatiotemporal features including operational conditions and environmental measures; (2) use deep sequential neural network models to capture the time-varying dynamic of go-around probability evolution; (3) analyze the features that contribute to high go-around probabilities for both the whole population and individual flights; and (4) develop a web-based user interface to display the results of the go-around prediction service. The contribution of this work is two-fold. First, we hope that with the help of the prediction service, both pilots and air traffic controllers can be alerted when there is a high risk of a go-around. Second, the feature contribution analysis can inform the different parties about the leading causal and predictive factors for go-arounds and suggest measures to mitigate go-around risk.

---

[1] Kaijing Ding and Ke Liu contribute equally.

The rest of the paper is organized as follows. Section 2 describes our data sources and the study airport. In Section 3, we introduce the detailed methodologies of this study. Section 4 presents the model results and Section 5 displays the real-time deployment of the model. The final section offers the conclusions and discusses the future work.

## II. Data and study area: JFK

This study focuses on arrivals in the years of **2019**, **2022**, and **2023** at John F. Kennedy International Airport (JFK), for its multi-runway layout and complex operational conditions.

### A. Data Sources

We collected four datasets for the three calendar years of 2019, 2022, and 2023 for JFK airport.

- **Integrated Flight Format (IFF), Reduced Data summary (RD), and *Events dataset (EV)*** from NASA Sherlock Data Warehouse: IFF records provide flight tracks updated every 1-10 seconds depending on the distance between the aircraft and the airport. RD file is a one-line summary for each flight in the facility and EV file records the details of an aircraft event (e.g., landing and takeoff). We use IFF data to obtain the flight trajectories (longitude, latitude, and altitude), ground speed, RD for aircraft type and airline information; and EV data for flight runway threshold crossing time and surface movement.
- ***Airport Configuration Data (APTC)*** from National Traffic Management Log: it records runway configuration updates of all major airports, including the time of each runway configuration change, the new arrival and departure runway configuration, and aircraft arrival rates and departure rates.
- ***METeorological Aerodrome Reports (METAR)*** from Aviation Weather Center of NOAA: it provides hourly update of wind, visibility, ceiling, and meteorological conditions in the terminal area.
- ***N90 Go-Around Reports*** from ATAC corporation: it summarizes the go-arounds that occurred in the New York TRACON (N90). This dataset, although incomplete, served as a supplementary source for cross-validating our go-around detection algorithm.

### B. JFK airport

JFK International Airport, a key hub in New York City, was the sixth busiest in the U.S. in 2019, with over 460,000 flight operations [14]. The airport has two pairs of parallel runways, all equipped with Instrument Landing System (ILS), except for runway 13R. In the three study years, runway 22L was the primary arrival pathway, handling more than 1/3 of traffic, followed by 04R (24.6%) and 31R (15.9%) as shown in Table 1. Notably, runway 13L/31R underwent closure from April to November 2019 for major reconstruction [15], reducing its usage compared to handling more arrivals in other years.

To identify the go-around events occurred at JFK from trajectory data, we modified the go-around detection algorithm introduced in [11] by (1) adding another distance check which requires the maximum distance between the go-around flight and runway threshold after the proposed go-around initiation point should be greater than the distance between them at the point; and (2) fine-tuning the parameters to make sure that our algorithm detects each go-around reported in the N90 Go-Around Reports that is visually confirmed, using Google Earth, to be real go-arounds. As a result, 2,302 go-arounds are detected in JFK in the three years, accounting for 0.39% of all arrivals.

TABLE I.    JFK Runway Usage and Go-Around Occurrence Rate

| Runway | Usage Percentage in Terms of Arrival Operations | Go-around Occurrences per 100 Arrival Operations |
|---|---|---|
| 04L | 3.6% | 1.11 |
| 04R | 24.6% | 0.41 |
| 13L | 8.0% | 0.69 |
| 13R | 0.05% | 0.67 |
| 22L | 38.7% | 0.31 |
| 22R | 4.0% | 0.69 |
| 31L | 5.2% | 0.28 |
| 31R | 15.9% | 0.19 |

The rate of go-around occurrences significantly varies by runway, as detailed in Table 1. Despite being one of the least utilized for arrivals (3.6%), Runway 04L exhibits the highest go-around rate at 1.11 per 100 operations, with Runways 13L (0.69%), 22R (0.69%), and 13R (0.67%)—also less frequently used for arrivals—following closely. In contrast, the most used runways for arrivals, 04R, 22L, and 31R, report go-around rates around or below 0.40%. Several factors might explain these variations. Firstly, runway selection often hinges on weather conditions; less-used runways might be selected under adverse weather, which may impact approach stability and cause go-arounds. Second, the less-used arrival runways, when used as arrival runways, are usually used as mix-use runways that accommodate both arrival and departure flights at the same time, which may generate additional risk (e.g., simultaneous runway occupancy and failed separation) for safe landings and lead to go-arounds. Third, pilots may be less experienced to land on the less-used runways, which leads to more frequent unsuccessful landing attempts. We integrate these considerations in feature engineering section to assess their influence on go-around probabilities more comprehensively.

## III. Methodology

### A. Model specification

This section outlines our modeling method to predict sequential go-around probabilities along landing approach, involving data processing, feature engineering, model training with hyperparameter tuning, and performance evaluation.

TABLE II.　　Model Variables Description

| Group | Variable | Description |
|---|---|---|
| (i) Flight specific information | Type_[x]+ | 1 if flight is operated by an international airline, 0 otherwise |
| | WC_[x]+ | Dummy variable for aircraft weight class |
| | Body_[x]+ | 1 if flight is wide-body aircraft, 0 otherwise |
| | month_[x]+ | Dummy variable for flight month |
| | dow_[x]+ | Dummy variable for flight day of the week |
| | tod_[x]+ | Dummy variable for flight time of the day |
| (ii) Final Approach Stability | groundspeed | Flight ground speed, (+/-) in knots |
| | energy | Kinetic energy height, (+/-) in feet |
| | horiz | Horizontal deviation from extended runway centerline, in meters |
| | alt_dev | Altitude deviation from 3° glideslope from revised landing threshold, (+/-) in 100 feet |
| (iii) In-trail Relation | lead_ent_D | 1 if leading aircraft is a departure, 0 otherwise |
| | lead_off_rwy | 1 if leading aircraft has got off runway, 0 otherwise |
| | separation | Separation from leading aircraft, in nautical miles |
| | hat_ROB | Predicted runway occupancy buffer, in seconds |
| | speed_diff | Ground speed difference from leading aircraft, in knots |
| | alt_diff | Altitude difference from leading aircraft, in 100 feet |
| | closing | 1 if separation from leading aircraft is decreasing, 0 otherwise |
| | lead_rwy_time | 0 if leading aircraft has not yet landed, otherwise runway occupancy time of leading aircraft, in seconds |
| | WC_trail | Weight class of trailing aircraft |
| | WC_lead_arr | Weight class of leading arrival aircraft |
| (iv) Airport/ Runway Conditions | rwy_[x]+ | Dummy variable for landing runway |
| | mix_rwy | 1 if landing runway is also used for departure when the aircraft is landing, 0 otherwise |
| | crs_rwy | 1 if the landing runway intersects with other runways in use, otherwise |
| | dep_rwy | 1 if the arrival runway configuration is 22L/22R, 4R/4L or staggered 31L/31R when the aircraft is landing, 0 otherwise |
| | ADR | Airport departure rate |
| | AAR | Airport arrival rate |
| | arr_ratio | Ratio between number of flights whose altitude is below 200 FL in TRACON and airport arrival rate |
| | #Obj_rwy | Number of objects on landing runway |
| (v) Weather | weather_[x]+ | Dummy variable for meteorological condition |
| | visibility | Airport visibility, in miles |
| | ceiling | Airport ceiling condition, in 100 feet |
| | head | Head wind, in knots |
| | tail | Tail wind, in knots |
| | cross | Cross wind, in knots |
| | gust | Gust wind, in knots |

+ Variables are one-hot encoded.

### 1) Flight sequences and features

We apply linear interpolation to the raw flight tracks, which update every 1-10 seconds, to create subsampled points every 0.5 nautical mile (nm) within 10 nm approach to the landing runway threshold. This generates a maximum of 20 data points per flight, ending at the start of a go-around initiation or at the runway threshold for normal landings. Each point, termed a cutoff gate, ranges from gate 0 (at 10 nm out) to gate 19 (at runway threshold). Go-around flights are assigned with a consistent binary label of 1 up to the go-around initiation gate; subsequent points are excluded. For instance, a flight initiating a go-around at 4.4 nm will have all points from 10 to 4.5 nm labeled as 1. Points beyond this, from 4 to 0.5 nm, are omitted. The result is a dataset with up to 20 labeled gates per flight, each gate featuring a compiled vector of attributes and a go-around indicator.

Based on prior research and data availability in Section 2, we derive five groups of features at each cutoff gate, as summarized in Table 2; and they can be further categorized into either static or dynamic features. Static features, such as flight specific information and airport configurations, remain invariant during the final approach. On the contrary, dynamic features such as altitude and speed will change through the approach phase. The entire dataset undergoes a Train-Validation-Test split (70%, 20%, 10%) for model training, early-stopping validation, and testing, respectively, with a pre-split random shuffle to ensure representativeness. Numerical features are standardized by gate based on training set statistics, which same statistics are then applied to the validation and test sets to maintain consistency. More details on the feature engineering process are available in the [11] and [12].

### 2) Model architecture and training

Utilizing trajectory data with environmental and airport operational conditions, our goal is to convert the multivariate time series of approach phase into sequential go-around probability predictions. Hence, we employ a Long Short-Term Memory (LSTM) network, a type of recurrent neural network designed to capture temporal dependencies in sequential data.

To accommodate the variable lengths of flight sequence of go-around flights and normal flights, we implement padding to streamline the input data, facilitating the encoding of sequences into contiguous batches for the LSTM network. The network features an input layer shaped to process the temporal sequence data, followed by a masking layer designed to manage the instances of early go-arounds. For flights undergoing a go-around prior to the terminal gate, all ensuing data points are masked, ensuring that model predictions rely solely on pertinent pre-go-around data, thus preserving the temporal sequence integrity and reflecting the constraints of operational practice.

At the core of the model, LSTM units are deployed to navigate the temporal dynamics of the aircraft's approach and landing phase. A dropout layer is added to prevent overfitting by randomly deactivating neurons during training iterations. Subsequent layers include a time-distributed dense layer, which fosters a robust connection at each timestep; and finally, an output layer with a sigmoid activation function that computes the likelihood of a go-around at each gate.

In the training phase, we choose Adam optimizer for its adaptive learning rate capabilities, and an early stopping protocol to optimize convergence and prevent overfitting. Given the extreme class imbalance in our dataset, with a minimal go-arounds fraction (0.39%), we adopt the binary focal cross-entropy loss (BFL) function, which is designed to enhance the model sensitivity to the underrepresented class [16]. The BFL function is particularly advantageous for our problem as it integrates a focusing parameter ($\gamma$) and a balancing term ($\alpha$), effectively accentuating the correct classification of the infrequent yet critical go-around events. It is computed as:

$$L(y,\hat{y}) = -\alpha y(1-\hat{y})^\gamma \log(\hat{y}) - (1-y)\hat{y}^\gamma \log(1-\hat{y}) \quad (1)$$

where $y \in \{0,1\}$ is a true go-around label; $\hat{y} \in [0,1]$ is an estimate of go-around probability; $\alpha$ is set at 0.95 to underscore the importance of the minority class; $\gamma$ is set at 2, echoing the recommendations from the focal loss literature [16].

To obtain the optimal model configuration, we employ a grid-search to tune hyperparameter. This process iterates through an expansive pool of 1,296 candidate permutations, derived from an array of LSTM units (128, 256), dense units (64, 128, 256), batch sizes (64, 128, 256, 512), activation functions (ReLU, tanh, swish), learning rates (0.01, 0.001), and dropout rates (0.0, 0.1, 0.2) to regularize the network. In addition, the loss functions within this tuning matrix encompass the standard binary cross entropy and the binary focal cross entropy, with an alpha value of 0.95 to account for class imbalance. Training incorporated early stopping callbacks to curb overfitting, while precision, recall, and AUC metrics gauged model performance. Each model iteration, along with its history, was preserved for analysis.

*3)   Model evaluation*

To evaluate the LSTM models for go-around prediction, we first use the test set to identify the optimal probability threshold for each model candidate. Each candidate model generates a probability $\hat{P}_{ij}$ indicating the go-around probability from gate $j$ to gate 20 at runway threshold for each test flight $i$. A flight with $\hat{P}_{ij}$ higher than the picked threshold will be viewed as a predicted go-around, and lower than that as a predicted non-go-around. We employ a grid search with a range of 0.1 to 1.0 to select the threshold that maximizes the F2 score (2) at each gate. This fine-tuning ensures the model sensitivity to be finely calibrated in detecting go-arounds at each gate along the approach path.

We then assess model performance using five key metrics: AUC-ROC, precision, recall, accuracy, and F2 score. Finally, the optimal model is selected based on the highest mean F2 scores across all gates. As the cost of missing a go-around detection highly outweighs the cost of getting a false alarm, we select F2 score for its weighted emphasis on recall, prioritizing the detection of actual go-around instances over avoiding false alarms, computed as (2).

$$F2 = \frac{(1+2^2) \times precision \times recall}{2^2 \times precision + recall} \quad (2)$$

where $precision = \frac{TP}{TP+FP}$ and $recall = \frac{TP}{TP+FN}$ with TP: true positive, FP: false positive, FN: false negative.

*B.   Model Explainability*

This section present method to identify key factors behind go-around decisions through two approaches: global analysis for identifying main predictors and local analysis for examining specific flight influences and potential causal relationships.

*1)   Global interpretability*

we conducted a permutation-based importance analysis on the test dataset to pinpoint dynamic features—those varying during flight arrival as detailed in Section 3.1.1.—that significantly affect go-around probability, indicating strong predictive power within the model. Our methodology, outlined in Table 3, emphasizes conducting N (e.g., 20) iterations for each feature at each gate to counteract random data fluctuations and solidify confidence in the identified feature weights [17, 18]. The BFL metric, chosen for its sharp sensitivity to class imbalances, serves as criterion for assessing feature importance, focusing specifically on dynamic features crucial for predicting go-arounds.

TABLE III.   ALGORITHM ON FEATURE IMPORTANCE USING PERMUTATION

**INPUT**: Trained LSTM model (m), test set features (X), go-around labels (y)
Compute original model performance using Binary Focal Loss (BFL):
   Ori = BFL(y, m(X))
for each cutoff gate j in {0, ..., 19} do
     for each dynamic feature i in X do
        Initialize an empty list to store importance scores: IptScores = []
        for iteration in {1, ..., N} do
           Randomly shuffle the values of feature i at gate j to create a perturbed feature matrix X_perm
           Compute the BFL on the perturbed dataset:
              Perm = BFL(y, m(X_perm))
           Calculate the importance score for the current iteration:
              PI$_{ij}$ = Ori - Perm
           Append PI$_{ij}$ to IptScores
        end for loop
        Compute the average importance score for feature i at gate j:
           β$_{ij}$ = Average(IptScores)
     end for loop
end for loop
**OUTPUT**: A matrix of averaged importance scores β, where each entry β$_{ij}$ represents the importance of feature i at gate j.

*2)   Local interpretability*

Given the global feature importance weights ($\boldsymbol{\beta_{ij}}$), where $i$ indexes the feature and $j$ indexes the cutoff gate, the relative contribution ($\boldsymbol{RC_{ij}}$) of each feature for a given flight at a particular gate is computed as follows,

$$RC_{ij} = f_i(\delta_{ij}) \times max(0, \beta_{ij}) \qquad (3)$$

where $f_i(\delta_{ij})$ is the feature-specific treatment function and $\delta_{ij}$ is the feature deviation (i.e. $x_{ij}$) after data standardization. The treatment function is customized to capture both the magnitude and directional impact of the feature's deviation from the mean. The function is defined as follows:

$$f_i(\delta_{ij}) = \begin{cases} 1_{x_{ij}}, & x_{ij} \in \{lead\_off\_rwy, closing\} \\ |\delta_{ij}|, & x_{ij} \in \{groundspeed, energy\} \\ |min(0, \delta_{ij})|, & x_{ij} \in \{visibility, ceiling, head, lead\_rwy\_time, \\ & \quad alt\_diff, separation, ADR, AAR, hat\_ROB\} \\ max(0, \delta_{ij}), & x_{ij} \in \{tail, cross, gust, speed\_diff, arr\_ratio\} \\ \left(\delta_{ij} - \frac{0 - \mu_{ij}}{\sigma_{ij}}\right), & x_{ij} \in \{horiz\_dev, alt\_dev, \#obj\_rwy\} \end{cases} \qquad (4)$$

Binary variables, **lead_off_rwy** and **closing**, simply indicate the condition's presence or absence. For features where any deviation is impactful, such as **groundspeed** and **energy**, the absolute value of the deviation is taken. For features of which higher values indicate less risk (**visibility**, **ceiling**, **head**, etc.), the negative deviation is used, and conversely, for features of which lower values indicate less risk (**tail**, **cross**, **gust**, etc.), the positive deviation is used. Deviations for features like **horiz_dev**, **alt_dev**, and **#obj_rwy** are adjusted by their standardized mean and standard deviation to track specific deviations from zero. Conditional features add another layer of analysis complexity. For flights with **lead_ent_D** or **lead_off_rwy** is true, any contribution from features such as **alt_diff** and **separation** are nullified, reflecting their diminished relevance when following another aircraft.

The contributions are then normalized to yield an alpha score ($\alpha_{ij}$, in percentage), enabling a comparative analysis across all features and gates, as (5). This normalization factor ensures that the sum of all contributions for a gate equates to 100%, facilitating a straightforward interpretation of factor dominance.

$$\alpha_{ij} = \frac{RC_{ij}}{\sum_{k \in X_j} RC_{kj}} \times 100\% \qquad (5)$$

IV. MODEL RESULTS

A. *Model training and performance*

This section discusses the results of training and validating the LSTM model for go-around detection. After data cleaning, there are **585,680** flights and **2,302** go-arounds in the final data set. A data split yields a training set with 1,611 go-arounds, a validation set with 460 go-arounds, and a test set with 231 go-arounds. The model, selected from 1,296 candidates via grid search for its best average F2 score of 0.7922, features 128 LSTM units and a dense layer with 256 units. The 'ReLU' activation function, with a Binary Focal Loss (BFL) function weighted with an alpha of 0.95, improves the model ability to predict sparse go-around events.

Table 4 reports the testing set performance of the final model. The "omitted go-around" column indicates the number of go-around flights omitted after they have been initiated, which ensures the performance metrics, specifically recall and precision, not to be inflated by subsequent timestamps after a go-around event has already been detected. The probability thresholds, varying from 0.47 to 0.63 across different gates, reflect the adaptive strategy to capture the varying likelihoods of go-around events. A recall (i.e., $\frac{TP}{TP+FN}$) value between 0.89 and 0.93 indicates that our model can predict approximately one out of three to five go-arounds. Precision (i.e., $\frac{TP}{TP+FP}$) values demonstrate that 34% to 60% of our predicted go-arounds are true. This may be explained by the low industry conformance rate of go-around policies that pilots do not always initiate a go-around even when the flight approach is unstable [5].

TABLE IV. MODEL PERFORMANCE

| Gate | Omitted Go-arounds | Threshold | F2 | Precision | Recall | Accuracy |
|---|---|---|---|---|---|---|
| 0 | 0 | 0.626 | 0.8243 | 0.5856 | 0.9177 | 0.9971 |
| 1 | 0 | 0.586 | 0.8205 | 0.5695 | 0.9221 | 0.9969 |
| 2 | 2 | 0.615 | 0.8181 | 0.6006 | 0.8996 | 0.9973 |
| 3 | 4 | 0.561 | 0.8177 | 0.5649 | 0.9207 | 0.9969 |
| 4 | 5 | 0.607 | 0.8126 | 0.5959 | 0.8938 | 0.9972 |
| 5 | 9 | 0.586 | 0.8152 | 0.5755 | 0.9099 | 0.9971 |
| 6 | 11 | 0.601 | 0.8162 | 0.5870 | 0.9045 | 0.9972 |
| 7 | 15 | 0.610 | 0.8116 | 0.5938 | 0.8935 | 0.9973 |
| 8 | 19 | 0.600 | 0.8077 | 0.5870 | 0.8915 | 0.9973 |
| 9 | 24 | 0.617 | 0.8090 | 0.6040 | 0.8841 | 0.9975 |
| 10 | 35 | 0.572 | 0.8082 | 0.5691 | 0.9031 | 0.9974 |
| 11 | 48 | 0.575 | 0.8105 | 0.5685 | 0.9071 | 0.9976 |
| 12 | 58 | 0.542 | 0.8071 | 0.5427 | 0.9191 | 0.9975 |
| 13 | 74 | 0.532 | 0.7901 | 0.5162 | 0.9108 | 0.9975 |
| 14 | 93 | 0.508 | 0.7702 | 0.4737 | 0.9130 | 0.9974 |
| 15 | 113 | 0.569 | 0.7489 | 0.4880 | 0.8644 | 0.9979 |
| 16 | 130 | 0.485 | 0.7346 | 0.4061 | 0.9208 | 0.9975 |
| 17 | 145 | 0.482 | 0.7286 | 0.3902 | 0.9302 | 0.9978 |
| 18 | 164 | 0.470 | 0.6858 | 0.3370 | 0.9254 | 0.9978 |
| 19 | 0 | 0.626 | 0.8243 | 0.5856 | 0.9177 | 0.9971 |

B. *Model Explainability*

This section presents interpretability of our LSTM network, including global and local factor contribution, to identify key predictors of go-arounds and how specific flight conditions affect predictions in real time.

In the global analysis of our LSTM model, we implement Permutation Importance to assess the impact of dynamic features across different cutoff gates, utilizing Binary Focal Loss (BFL) as our evaluative metric for its robustness against class imbalance prevalent in our dataset. To ensure the stability of our findings, we have iterated the permutation process for each feature at each gate 5 times. The resulting measures of feature importance are then visualized as barplots in Fig. 1(a), which delineates the top 10 features for each specific gate. The x-axis of these plots lists the feature names, while the y-axis quantifies their importance based on the observed variation in BFL from the permutations. Noteworthy within these plots is the prominence of *lead_rwy_time*, *#Obj_rwy* and *energy_pos* as the most significant predictors in the model. Conversely, we observe that certain features, like *groundspeed*, exhibit minimal and even negative importance at specific gates, hinting at potential overfitting or noise with these features.

For local interpretability, we present flight BWA550's go-around at JFK on Aug. 8th, 2019, to demonstrate how feature variations affect go-around probability predictions (analyzed based on a previous version of model candidate). As showed in Fig. 2, BWA550 initially aimed for runway 22L but had to go around at 4.93 nm from the threshold, later landing on 22R. Based on the LSTM model and feature contribution analysis, we dissect the factors affecting the go-around probability as the aircraft approached from 10nm to 6.5nm. In Fig.1(b), the go-around probability varies between 0.457 and 0.787, indicating a dynamic assessment of the flight conditions. Predominantly, factors like *hat_ROB* and *lead_rwy_time* significantly influenced the predictions, making up over 80% of the contributory impact. The significant weights assigned to these factors suggest that the model perceives a high risk of simultaneous runway occupancy with its leading arrival flight, which is highly likely to lead to a go-around decision.

The result displays the ability of our prediction model in real-time risk assessment and its usefulness as a decision-support tool in landing phase. The feature contribution analysis offers explanations for the predicted go-around probabilities, which can potentially help the pilots and air traffic controllers develop measures to mitigate the go-around risks.

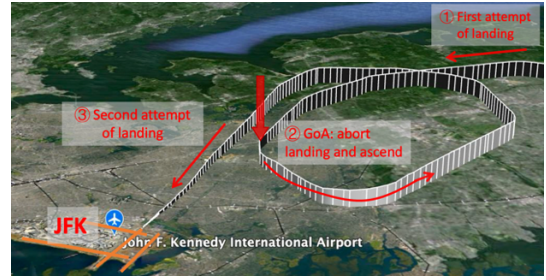

Figure 2. Approaching Path of Flight BWA550 at JFK

## V. REAL-TIME DEPLOYMENT

### A. Overview

One of the novelties of this study is its ability to be deployed to real-time streaming services. As illustrated in Fig. 3, our workflow updates every 5 seconds to provide go-around probability predictions. After collecting IFF, RD and EV data for 5 seconds, we update the latest APTC and METAR data for runway and weather condition, respectively. For each arriving flight updated within this timeframe, we (1) verify it's within 12.5 nm of the airport center, under 5,000 ft altitude, AND not landed—else, we move to the next; (2) predict targeted runway and its leading and trailing flight (if any) using the algorithm in the next section; (3) prepare feature matrix according to Section 3.1; (4) predict go-around probabilities with the pre-trained LSTM model; (5) output go-around probabilities with top contributing features, and other variables of interest.

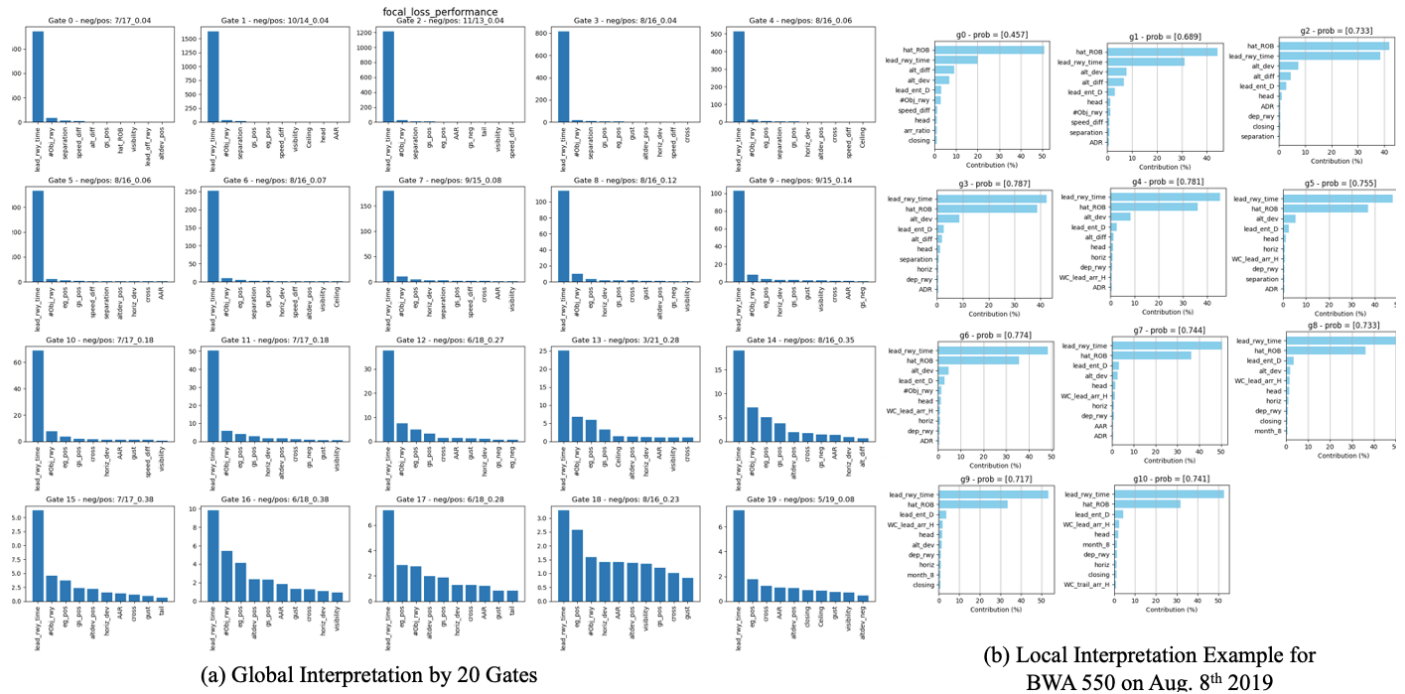

(a) Global Interpretation by 20 Gates

(b) Local Interpretation Example for BWA 550 on Aug. 8th 2019

Figure 1. Interpretation on go-around events via LSTM network on global and local scope

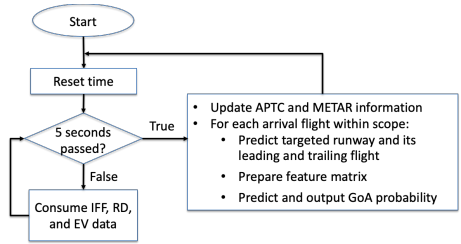

Figure 3.   Flow Chart of Real-Time Prediction Service

## B. Real-time Runway Prediction

In a real-time setting, unlike with historical data where the actual landing runway is known, predicting the targeted runway for an incoming flight is challenging due to the lack of direct runway assignment information. This prediction is inferred from the flight trajectory and current runway configurations, which are typically limited to a maximum of two runways (shown in in Table 1) for arrivals at any given time at JFK. Leveraging real-time data on runway configurations from APTC, we can make a random "guess" on the intended runway, with a baseline accuracy of at least 50% for our predictions.

For enhanced accuracy, we scrutinize the flight paths as depicted in Fig. 4, where structured arrival routes align with extended runway centerlines during final approaches. This pattern implies the feasibility of predicting runways by calculating flights' perpendicular distance to each runway's extended centerline and picking the one with the shortest distance. However, challenges arise due to the final approach span being under 10 nm, especially runway 13L with final approach of merely 4.5 nm. Overlaps in approach paths (see the red dots in Fig. 2) indicates that it would be impossible to reach 100% accuracy using flight positions alone.

We refine our method by incorporating flight course data, particularly for the 13L's short final approach. Note that 13L/22L is the only dual runway operation of 13L. As 13L serves arrivals from the south and 22L from the north, we predict landings on 13L for flight courses under 90° and on 22L for courses between 180° and 270°. Additionally, we add heading and location checks to validate runway predictions. The heading check confirms flights are directed towards the runway start, decreasing in distance, while the location check ensures flights are within a set perpendicular distance from the runway centerline—8.5 nm for runway 13L, 4 nm for others—AND closer to the runway beginning than runway end. These checks validate proximity to final approach, facilitating runway-specific sequencing of arrivals by proximity to the threshold and determination of leading and trailing aircraft.

The complete algorithm is described in Table 5. We test the runway prediction of all different runway configurations with real-time flight data spanning ten days in 2019. Fig. 5 show the prediction results. The blue curve represents the proportion of flights with runway predictions at varying distances from runway threshold, while the orange curve shows the accuracy of these predictions. The analysis can extend up to 15.5 nm from runway threshold, though our real-time analysis begins within a 12 nm radius of the airport geometric center. Notably, prediction accuracy is initially low but improves significantly from 12 nm to 10 nm, with around 90% of flights receiving accurate predictions at the 10 nm mark. The accuracy discrepancy between the two curves lessens from 10 nm to 8 nm, a range where overlapped approach paths for parallel runways occur (where the red dots in Fig. 4 fall in). Below 8 nm, prediction accuracy for both curves approaches 100%. Although current accuracy is deemed satisfactory, future efforts may explore advanced probabilistic methods to refine these predictions.

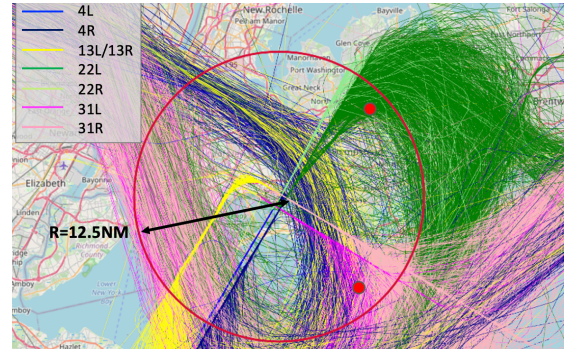

Figure 4.   Trajectories of Flights Arriving at JFK on 1/1-1/10/2019

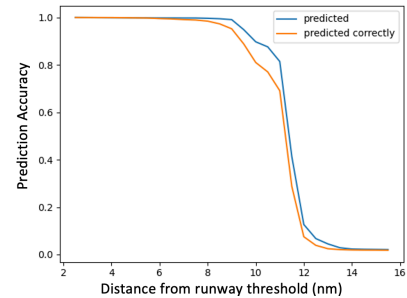

Figure 5.   Accuracy of the Real-Time Runway Prediction Algorithm

TABLE V.   REAL-TIME RUNWAY AND IN-TRAIL PREDICTION ALGORITHM

**INPUT:** 4D flight track data (latitude, longitude, altitude, and time)
**INITIALIZE:**
- Coordinates of all runway beginnings and runway ends at the airport.
- An empty nested dictionary, rwy_sequence, to track flight approach sequences for all runways, mapping each runway to its arrivals with its arrival time, distance to runway beginning, and latest ground speed.

**OUTPUT:** Predicted runway, lead flight ID (if any), trail flight ID (if any)

**Step 1: Predict runway based on APTC runway configuration.**
- If only one arrival runway is in use, predict it.
- If more than one arrival runway is in use:
    a. For APTC configurations 13L/22L, if the flight course is under 90° or between 180° and 270°, predict 13L for courses under 90°, and 22L for the rest.
    b. Otherwise, predict the runway closest in perpendicular distance to each candidate's extended centerline.

**Step 2: Heading check.** If the flight is flying towards the predicted runway beginning, continue to next step; otherwise, stop prediction for this flight.

**Step 3: Location check.** If the flight is within certain perpendicular distance of predicted runway centerline AND is closer to runway beginning than runway end, continue to next step; otherwise, stop prediction for this flight.

**Step 4: Update rwy_sequence.** Sort flights by their straight-line distance to the runway's start at the current time step.
**Step 5: Identify the leading and trailing flight based on the sorting result.**
**end procedure**

## C.　User Interface

This section describes the web-based interface for our go-around prediction service, currently under development. Fig. 6 shows a screenshot of its latest version, showcasing real-time positions of flights near JFK terminal on Feb. 2nd, 2024. The left window shows arrivals queued to land on runway 31L/31R, while right side lists the selected (circled) flight DAL 1547 with feature values and go-around probability of 11% at that moment.

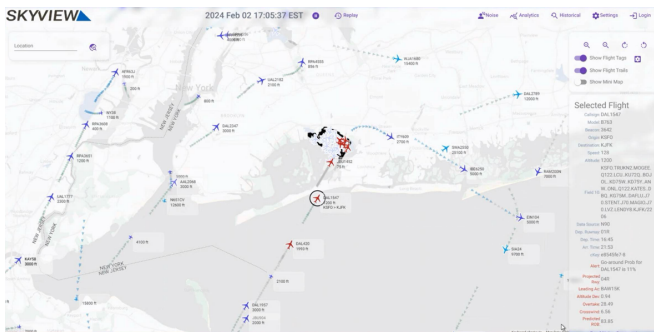

Figure 6.　Screenshot of the Web-Based User Interface

## VI.　Conclusion and Discussion

In this paper, we employ the LSTM network to predict the real-time go-around probability as a flight is approaching the airport within 10 nm of the landing runway threshold. LSTM model is chosen for its ability to capture the time dependency in the flight sequence and provide an evolutionary and consistent go-around probability. We further develop methods to analyze the feature contribution to go-around occurrence both from a global view and an individual flight perspective. From our results, runway occupancy buffer, runway occupancy time of the leading aircraft, as well as separation with the leading aircraft, appear to be the factors that contribute the most to overall go-around occurrences. We then integrate these pre-trained models and analyses with real-time data streaming, and finally develop a demo web-based user interface that integrates the different components designed previously into a real-time tool that can eventually be used by flight crews and other line personnel to identify situations in which there is a high risk of a go-around.

Future work can enhance our predictive model by expanding the feature space with detailed datasets, incorporating granular aircraft dynamics (e.g., deceleration rates, descent rates, heading degrees) and weather conditions (e.g., wind shear). Improving data quality and aligning feature engineering with aviation standards can also increase model accuracy. Structurally, introducing gate-specific weighted dense layers could refine temporal analysis of flight approaches, and adding more dense layers may deepen feature interaction insights. The go-around detection algorithm and prediction model has been adapted for other airports like SFO with minimal modification and they perform well with similar accuracy. Further avenues for development include expanding the dataset with additional years of data and adapting the models for more airports.